# A SIMPLIFIED APPROACH FOR QUALITY MANAGEMENT IN DATA WAREHOUSE

Vinay Kumar[1] and Reema Thareja[2*]

[1]Professor, Department of IT, VIPS, GGSIPU, New Delhi 110 088, India
[2]Assistant Professor SPM Delhi University,
Author of Oxford University Press, India
*Corresponding Author

### ABSTRACT

*Data warehousing is continuously gaining importance as organizations are realizing the benefits of decision oriented data bases. However, the stumbling block to this rapid development is data quality issues at various stages of data warehousing. Quality can be defined as a measure of excellence or a state free from defects. Users appreciate quality products and available literature suggests that many organization`s have significant data quality problems that have substantial social and economic impacts. A metadata based quality system is introduced to manage quality of data in data warehouse. The approach is used to analyze the quality of data warehouse system by checking the expected value of quality parameters with that of actual values. The proposed approach is supported with a metadata framework that can store additional information to analyze the quality parameters, whenever required.*

## 1. INTRODUCTION

Many business organizations, large or small, are implementing data warehouses to collect, store, and process large amount of data. This enables organization to optimize the costs associated with data and in making smart and analytical decision. It further facilitates management of the organization to devise strategic policy to harness optimal returns on investment [1]. However, the size and complexity of data warehouse systems [3] make data prone to error that compromises quality.

The success of data warehousing initiatives depends on the quality of the data stored in it [2]. Research and industry surveys indicate that organisations experiencing problems with data quality are on constant rise [3]. This calls for a serious approach to manage data quality. Although data quality issues have a direct economic and social impact [4] yet little work is done for formulating a framework for measuring, evaluating, and improving data quality [5].

Good quality data ensures user's trust in data warehouse system making it more usable and, optimizes the business benefits gained. However, detecting defects and improving data quality comes with a cost and if the targeted quality level is high, the costs often negate (offset) the benefits. Given the economic trade-offs in achieving and sustaining high data quality in data warehouse, a framework of activities are used to measure quality of data in the data warehouse. In this paper, we propose a simplified approach for managing data quality in data warehouse systems. A block diagram of conceptual framework for data quality measurement and its implementation is shown in the Figure 1.

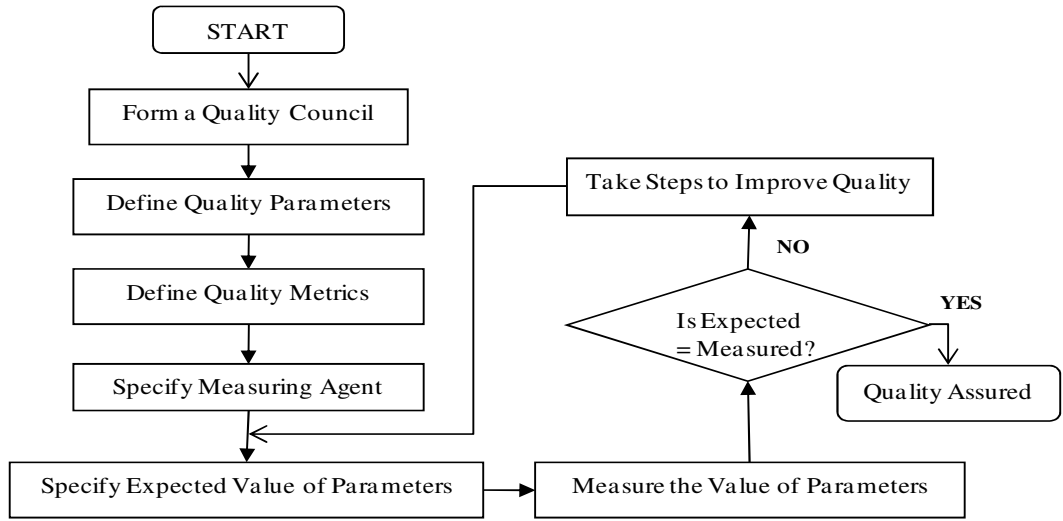

Figure 1: Conceptual framework for data quality measurement in data warehouse

TABLE 1: Possible data quality issues for stakeholders

| Stakeholder | Role | Data Quality Issues |
|---|---|---|
| Decision Makers | Final users who uses reporting tools, OLAP, Data mining to get answers to their questions | Overall quality, ease of access, reports in desired format and timeliness |
| Data Warehouse Administrator | Keeps data warehouse properly operating | Error reporting, timeliness and metadata accessibility |
| Data Warehouse Designer | Designs of data warehouse architecture | Schema design, metadata quality design and software quality design |
| Data Warehouse Programmer | Develops actual data warehouse applications | Implementation quality, overall software quality, metadata quality |
| Executive Manager | Concerned with financial information regarding data warehouse | Keeps a check on costs, benefits and return on investments |

The data warehouse development process must incorporate the data quality requirements of all potential stakeholders of the data warehouse environment. For this, the development team must understand the aspects of data quality that are important to each group. Different stakeholders may have different perspectives on data quality, so it is necessary to identify differing data quality requirements. Different stake holders of a data warehouse in an organization, their roles in data warehouse and possible data quality aspect [4] they are interested in, are organized in the Table 1. The paper is organized in six sections. Section 2 deals with source of data quality problems. A framework for measuring quality of data is discussed in the Section 3. In Section 4, criterion for improving upon data quality is discussed. Section 5 deals with meta data based quality model and the paper is concluded in the Section 6.

## 2. SOURCE OF DATA QUALITY COMPROMISE

For any data warehouse system, data flows from the operational systems, to the data staging area where the data collected from multiple sources is integrated, cleansed and transformed; and from staging area, data is loaded in to the data warehouse. Quality of data can be compromised depending upon how data is extracted, integrated, cleansed, transformed and loaded in the data warehouse. These stages are source of error or stages of possible quality compromise. All these stages need to be monitored to find any defects or discrepancies in the stored data. This is done by making a list of quality parameters, defining the metrics for identified each parameter and then comparing the specified metric value with that of actual value.

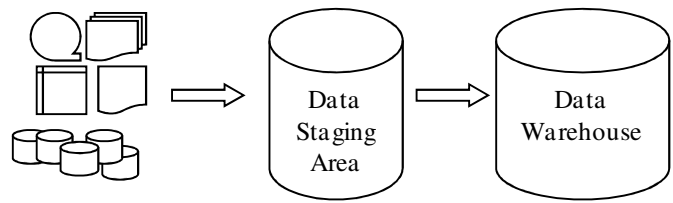

Figure 2: Information flow

Data quality compromised in one stage leads to errors in the next successive stages. Study indicates that following are the potential reasons for compromise in quality of data at different stages of data warehousing [6].

a. Selection of source systems that do not comply with business rules
b. Lack of data validation techniques practiced in source systems
c. Representation of data in different formats
d. Inability to update data in a timely manner
e. Data inconsistency problems in source systems
f. Lack of data quality testing performed on source systems
g. Missing values in certain columns of source system's table
h. Different default values used for missing columns in the source system
i. Disabling data integrity constraints in data staging area.
j. Absence of a centralized metadata component.
k. Ignoring the storage of data cleaning rules in the metadata repository.
l. Incomplete or wrong requirement analysis results in a poor schema design and hence another source of possible quality problems.

## 3. PROPOSED QUALITY MEASUREMENT FRAMEWORK

Quality data is always critical for the success of any data warehouse system. It helps in maintaining sustained competitive advantage and better customer relationships. It provides a new source of savings and also in making better organizational strategy [7]. However, datasets often suffer from defects like missing values, invalid entries, inaccurate data, and obsolete information [4]. Such type of erroneous or ill-quality data lacks customer satisfaction, hinders decision making process, increases costs, breeds mistrust towards data warehousing technology and has a negative impact on business performance [8]. The extent of the presence of quality defects in a dataset is measured as the ratio of the number of non defective records ($ndR$) and the total number of records (R). A framework of activities that is used to measure quality of data in the data warehouse is shown in the Figure 1.

$$Qulaity\_Defects = ndR / R$$

**Quality Council**

The first step to manage quality in data warehouse systems is to form a Quality Council [14] that will be responsible for quality parameters identification, quantification, evaluation and monitoring of the quality related issues. The council deals with formulation of an effective quality policy and a quality system that establishes and maintains documented procedures to control and verify that data and information stored in data warehouse meets the specified quality measure. The tasks of quality council may include the followings:

- Establishment of procedures for enforcing and maintaining quality of data warehouse.
- Documentation of procedures and data that affect the quality of data warehouse.
- Regular verification and validation of the report and tracing back to source to ensure that specified quality is adhered to.
- Maintenance of quality records that demonstrates the achievement of the required quality.
- Verification of the effective and economical operation of the entire data warehouse quality system.

Data warehouse quality may be defined using parameters enlisted in Table 2. The table also contains corresponding metric used for quantification of the measured value of these parameters. For each quality parameter, measuring agent(s) should be specified. If the measuring agent(s) has not been specified then the quality council must determine the computation procedure for the actual values of the quality factors by using the metrics specified in Table 2.

In addition to this, for each quality parameter the acceptable/expected values should also be clearly specified. This is required for the objective assessment of the subjective quality goal. The measuring agents communicate with the components of the data warehouse to extract measurements. Whenever the user's requirements change, the stakeholder may re-define the quality metrics and measuring agents responsible to deliver the base quality data.

**Measure and Compare Expected Values with Actual Values**

The next step is to measure the actual values of quality parameters using the specified measuring agents. Once the values are evaluated, these are compared with that of acceptable or expected values which have already been specified. If actual value is within the domain of acceptable values then, data warehouse satisfies the specified quality constraints. However, if the actual value is not in the acceptable range then technique specified in the following section for improving data quality is used to ensure the quality parameter is achieved.

For example, a quality measurement to calculate completeness will count the number of records with incomplete values. Then it will find the percentage i.e. number of incomplete records divided by total number of records and then multiplied by 100. If expected value for this parameter for quality purpose is set to 40%, then the quality of data warehouse will be accepted if the calculated percentage is less than or equal to 40%. Otherwise the data undergoes quality improvement process.

**Quality Parameters and their Metrics**

TABLE 2: Data Warehouse Data Quality Parameters and Metrics

| S.No. | Quality Parameter | Quality Metric |
|---|---|---|
| 1. | Functionality | Number of modules that are not appropriate for the task |
| 2. | Reliability | Number of failures |
| 3. | Usability | Acceptance by users |
| 4. | Efficiency | Performance in terms of response time, processing time, etc |
| 5. | Maintainability | Man hours required to maintain and test the applications |
| 6. | Portability | Number of cases where applications failed to work on new environments |
| 7. | Accessibility | Number of NULL values stored (where they are not expected) |
| 8. | Accuracy | Number of records with accurate values |
| 9. | Consistency | Number of records violating constraints |
| 10. | Security | Number of modules that could not protect the system from unauthorized access |
| 11. | Compliance | Number of modules non-compliant with standards/ conventions/ regulations |
| 12. | Recoverability | Number of times the software was unable to re-establish its level of performance and recover the affected data |
| 13. | Analyzability | Man-hours required for diagnosing defects or failures |
| 14. | Changeability | Man-hours required for removing defects from the system |
| 15. | Testability | Man-hours required for validating the software product |
| 16. | Install ability | Man-hours required for installation of the software in a specified environment |
| 17. | Implementation Efficiency | Number of resources used for the development of software and percentage of used recourses with respect to the originally expected ones |
| 18. | System Availability | Number of cases when relevant information is not available |
| 19. | Currency | Number of pieces of information where transaction time though required was not present |
| 20. | Volatility | Number of pieces of information where valid time though required was not present |
| 21. | Completeness | Number of records with incomplete values |
| 22. | Credibility | Number of records with inaccurate values |
| 23. | Data Interpretability | Number of pieces of information that are not fully described or documented |

## 4. IMPROVING DATA QUALITY

Errors in the data can be detected by comparing data to a correct baseline (like real-world entities, predefined rules or calculations, a domain of feasible values, a validated dataset) or by checking for missing values and by examining time-stamps associated with data. Correction of values is a complex task which often includes multiple inputs, outputs, and processing stages [9]. For this, organizations may consider correcting defects manually or hiring agencies that specialize in data enhancement and cleansing. Error detection and correction can also be automated by the adoption of methods that optimize inspection in retrieval of data from data warehouse while generating new information [10], integrity rule-based systems [11], and software agents that detect quality violations [12]. Some ETL (Extraction, Transformation, and Loading) tools also support the automation of error detection and correction.

Though error detection and correction policies help in improving data quality, yet this does not eliminate root causes of defects or reduce their impact. For this data processes needs to be built from scratch or, existing processes can be re-designed to better manage quality and reduce errors by embedding controls in processes, supporting quality monitoring with metadata, and improving operational efficiency. The data cleansing actions can be broadly classified as:

- *Prevention:* Data defects is prevented or at least reduced during data acquisition by implementation of simple and user friendly data acquisition user interfaces, disallowing missing values, validating values against the domain of feasible values and enforcing integrity constraints.

- *Audit:* Quality defects can occur during data processing when integrating data or due to wrong calculations or changes in the real-world entity that the data describes. The auditing process looks at the data to fully understand content, structure, completeness, distribution of values, and related factors. This detailed knowledge helps the design and enforcement of data filtering and correction policies. Data auditing is done at multiple points in the data warehouse life cycle. For instance, when new source of data is added to the data warehouse or during each transformation step or even after the data is loaded in the data warehouse.
  The audit process may be implemented as a series of programs or as a set of queries executed on the data. The output is then reviewed with the users. In case of defects, data filtering or correction is done for improving the data quality. The process tests data against integrity and reports any violations. This way it also helps in determining the scope of integrity rules.

- *Filter***:** During the audit process, if any erroneous data is found, the filtering process removes the data. Data filtering is to remove either individual data elements or entire records or logically related sets of data. Individual element is removed by setting its value to NULL. Row level filtering removes an entire row from result set which is not appropriate for a data warehousing target. Filtering a logical data group removes related rows from multiple tables in such a way that new integrity problems e.g. referential integrity violation is not introduced. Filtering should be used with caution as it may lead to incomplete information in the data warehouse. Moreover, removing data from the result set does not necessarily improve data quality. Contrarily, it results in loosing information that exists in operational systems and limits the ability of the data warehouse to respond to some business questions. However, for an effective filtering process, it is necessary to understand *why* integrity violations have occurred, and *how* the target data will be used.

- *Correct***:** When audit process reveals data that violates integrity rules, and filtering does not seem to be a viable solution, then that data needs to be corrected. Correction is done when data values accurately reflect business realities. If data quality problems are due to source systems then these problems are fixed in the source systems. However, corrections in source is often a slow process and at times even impractical or impossible. In such cases, data correction is done in data warehouse.

Another issue of concern is whether defects in data are worth correcting or not as correction might be time consuming and costly especially when missing data has to be procured. So data correction can be avoided when the added value cannot justify the cost. Repairs may involve finding an alternate data source, deriving a value from other data, using a default value, etc. data correction and repair can be done using action research involving different field of studies of Computer science.

## 5. PROPOSED META DATA BASED QUALITY MODEL

The proposed quality metadata model is given in the Figure 3. Every stakeholder in the data warehouse has a *quality goal* to evaluate, improve or administer the quality of entire data warehouse, or a part of it. A *quality goal* is an abstract requirement, defined on DW object and documented for a *purpose* in which the *stakeholder* is interested in.

*Quality dimensions* are used to define abstractly different aspects of quality, as the stakeholder perceives it. Quality goal is mapped to one or more *quality queries* that determine whether a goal is achieved or not. Each quality query is dispatched to *quality metrics* that describes measurements of quality. A quality metric is defined on a specific data warehouse object. It specifies an interval of expected values within the domain. It incorporates actual value within the domain at a particular point of time given by timestamp. The actual values of quality metrics are measured by a simple software agent, which acts as measuring agent.

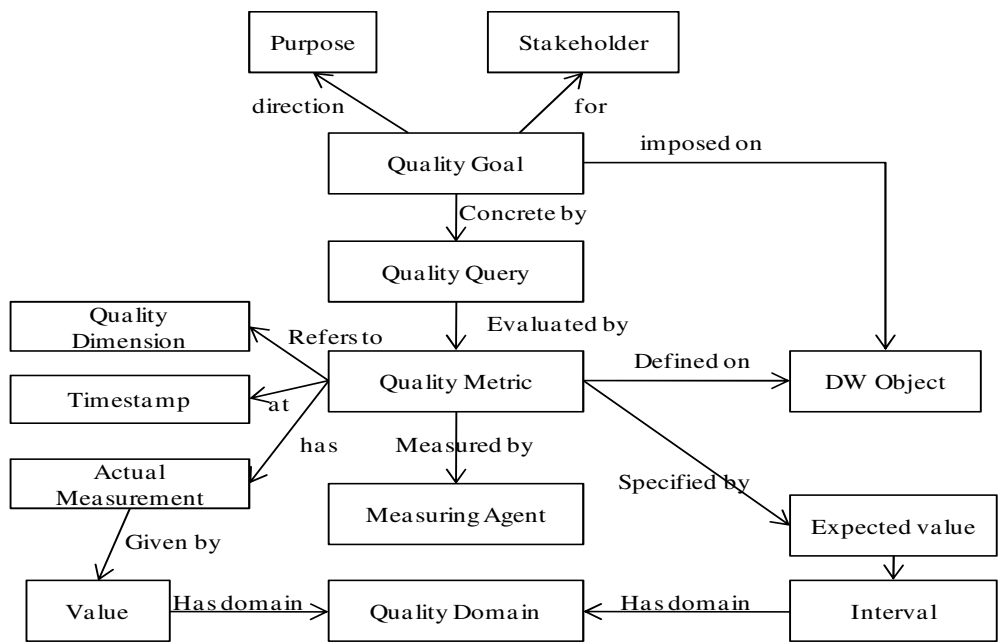

Figure 3: Proposed Quality Meta Model Framework

## 6. CONCLUSION

Data quality is an important factor in the success of data warehousing projects. We have identified potential source of errors causing quality compromises and presented in this paper. Meta data based quality model is proposed to enforce quality in the data warehouse. While evaluating the quality value of an object/component, if it fails to meet the specified quality level, then how to improve upon the quality of that component is also provided. It is not easy to quantify some abstract entity. Suitable metric to quantify identified parameters is also provided in this paper.

While implementing a data warehouse project in an organization, there are different stakeholders and they view data differently. And accordingly the importance of quality concept varies. While designing and implementing DW, it is important to understand and incorporate the expectation of all the stakeholders from the DW.

### ACKNOWLEDGMENT

We are grateful to all those who have been constantly encouraging us to go for such application oriented study work besides the regular work which we are doing at our respective departments.

**Biographical Notes:**

Vinay Kumar is a Professor in Vivekananda Institute of Professional Studies, Delhi. Earlier he worked as Scientist in National Informatics Centre, MoCIT, Government of India. He completed his Ph.D. in Computer Science from University of Delhi and MCA from JNU, Delhi. He has authored a book on Discrete Mathematics and has contributed many research papers to refereed journals and conferences. His area of interest is graph algorithm, steganography, data security, data mining and e-governance. He is member of CSI and ACM.

Reema Thareja is working as an Assistant Professor in Department of Computer Science, Shyama Prasad Mukherjee College for Women, University of Delhi. She is author of Programming in C, Data Strcutures, Data Warehousing, Data and File Structures (GTU), Computer Fundamentals and Programming in C, Introduction to C Programming and co-author Computer Programming and Data Structures (JNTU) all published by Oxford University Press.